\documentstyle[aps,twocolumn]{revtex}

\begin{document}

\wideabs{
\title{Comparative investigation of $^{39}$K and $^{40}$K trap loss rates:
alternative loss channel at low light intensities}
\author{G. Modugno, G. Roati\thanks{also at Dipartimento di Fisica, Universit\`a di Trento, 38050 Povo (Tn),
Italy}, and M. Inguscio}
\address{INFM and European Laboratory for Non-linear Spectroscopy\\
(LENS),\\
Universit\`{a} di Firenze, Largo E. Fermi 2, 50125\\
Firenze, Italy}
\author{M. S. Santos, G.D. Telles, L.G. Marcassa, and V.S. Bagnato}
\address{Instituto de F\'{\i}sica de S\~{a}o Carlos - USP\\
Caixa Postal 369, 13560-970, S\~{a}o Carlos-SP, Brazil}
\date{\today}
\maketitle

\begin{abstract}
We report a comparative investigation of trap loss rates in a magneto-optical trap for two
potassium isotopes, $^{39}$K and $^{40}$K, as a function of trap light intensity. 
The isotopes present a quite similar behavior for the loss rates at high intensities, 
and a sudden increase of the loss rates at low intensities is present in both cases. 
While for $^{39}$K such increase can be explained assuming that the major contribution to
 the losses comes from Hyperfine Changing Collisions, a different loss mechanism must be
 considered for $^{40}$K, which has an inverted ground state hyperfine structure. 
The experimental results of both isotopes are well reproduced by an alternative model 
based on radiative escape as the dominant loss mechanism.\\
\\
PACS number(s): 23.23.+x, 56.65.Dy
\end{abstract}

}

\section{Introduction}

The investigation of collisional losses in magneto-optical traps (MOTs) has
been extended to most of the trappable neutral atoms \cite{Weiner99}.
Although the modeling of trap losses has proved to be a nontrivial task, it
may reveal useful informations about the interatomic potentials and about
the nature of the inelastic processes, which constitute one of the main
limits towards the achievement of high densities in cold samples of atoms.

Since the first investigations of the trap loss behavior in alkali MOTs as a
function of laser intensity, the existence of two different loss regimes was
clear. While all later observations indicated that at large intensities the
losses are dominated by fine-structure-changing (FSC) and radiative escape
(RE), the origin of the loss behavior at low intensity is still not
completely understood. On one hand, the measurements on most of the alkali
species showed a sharp increase of loss rate for decreasing intensity below
a certain threshold value (usually, $I<30%
\mathop{\rm mW}%
/%
\mathop{\rm cm}%
^{2}$). Once all the studied species are trapped in their higher hyperfine
ground state in a MOT, the losses were attributed to hyperfine-changing
collisions (HCC), whose contribution increases as the trap depth decreases.
On the other hand, recent measurements on a $^{87}$Rb MOT showed that the
loss rate starts to decrease again at even lower intensities, while it would
be expected to raise away in presence of HCC \cite{Nesnidal}. These
experimental results are indeed well reproduced by an alternative model for
trap losses, assuming the RE as the dominant loss mechanism also for low
light intensity \cite{Telles}. In addition, the assumption for the presence
of trap loss channel other than HCC has been emphasized by the loss behavior
measured in a Cr MOT \cite{Bradley}. The overall behavior observed for the
Cr MOT trap loss rate is quite similar to those for the alkalis, including
the sharp rise up when the intensity decreases below some threshold value,
although the Cr atoms do not present hyperfine structure.

The fermionic potassium isotope, $^{40}$K, is particularly interesting in
this context, due to its inverted hyperfine structure for the ground
electronic state, unique among all stable alkali isotopes. This feature
should result in a suppression of HCC in a MOT, allowing better
investigation of trap loss mechanisms other than HCC. Actually, $^{40}$K has
been trapped in MOT only recently, due to its low natural abundance
(0.012\%) \cite{Cataliotti98,DeMarco99,Modugno99}, and to our knowledge no
trap loss measurement has been presented yet. On the contrary, the most
abundant bosonic isotope, $^{39}$K, presents a regular hyperfine structure.
And it had already been the subject of loss measurements in a MOT, and in
particular Santos {\it et al.} \cite{Santos98} have observed the
characteristic sharp increase for the trap losses at low intensity,
associated to HCC following the explanation given for other systems.

In this paper we present a comparative study of trap loss rates measured for
two potassium isotopes held in a MOT, $^{39}$K and $^{40}$K, as a function
of trap laser light intensity. The experiments were carried out in similar
experimental conditions, but in separate laboratories in Brazil and in
Italy. On the next section we present a brief description of the
experimental setup used during the trap loss measurements. Then we present
the experimental results, and we show that they are quite well reproduced by
a model based on RE \cite{Telles} as the main loss channel also at low light
intensity.

\section{Experimental setup}

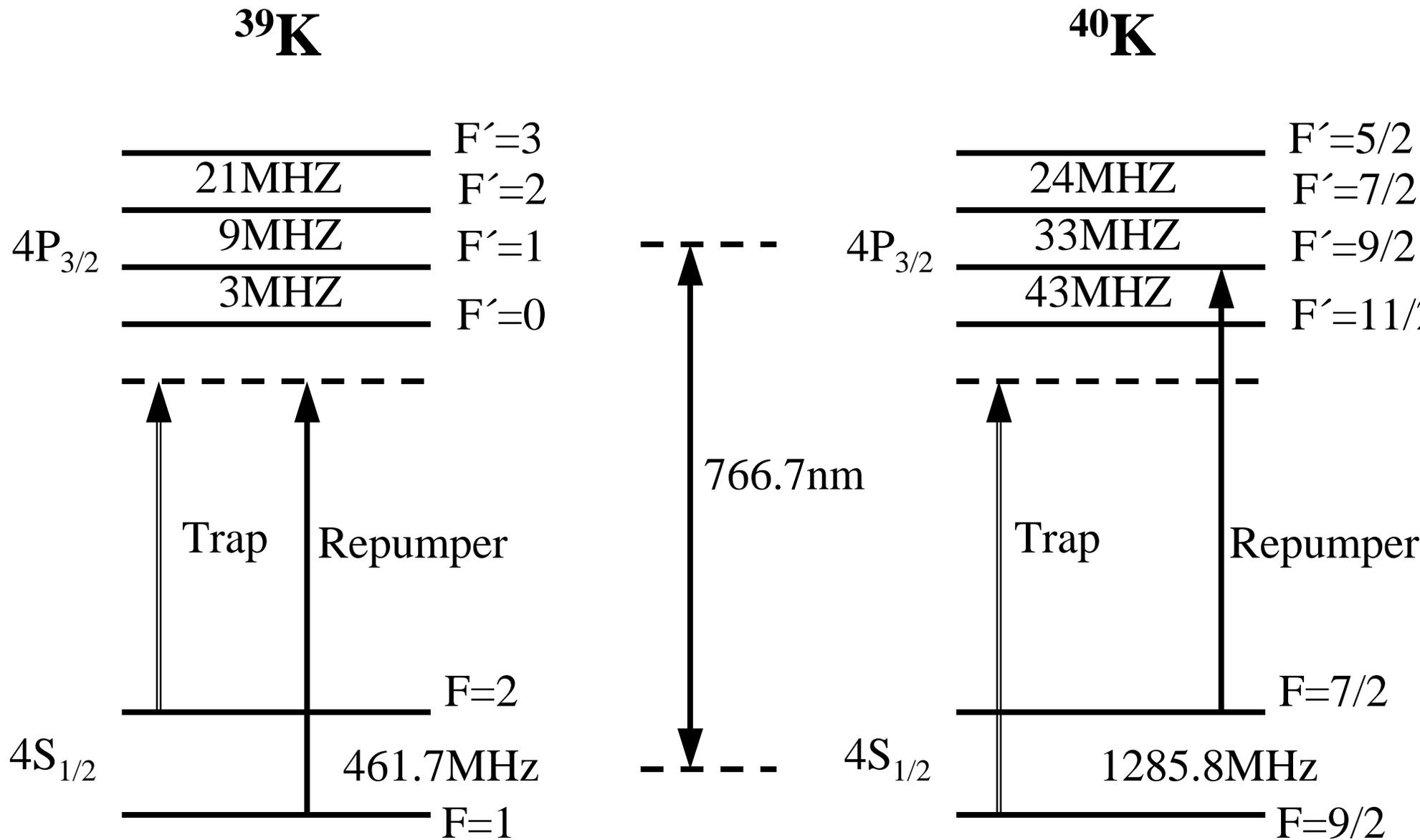
\begin{figure}[tbp]
\caption{Energy level scheme for the two isotopes $^{39}$K and $^{40}$K,
showing the hyperfine structure. Trap and repumper frequencies used for
magneto-optical trapping are indicated by arrows.}
\label{levels}
\end{figure}

The electronic structure of both isotopes investigated in this work are
shown in Fig.~\ref{levels}. For the bosonic isotope $^{39}$K, the ground
state (4S$_{1/2}$) presents two hyperfine levels with total angular momenta $%
F=2$ and $F=1$, frequency shifted by $462%
\mathop{\rm MHz}%
$. On the other hand, $^{40}$K presents a inverted hyperfine structure of
the ground state, with $F=7/2$ higher than $F=9/2$ by $1286%
\mathop{\rm MHz}%
$. Also in the excited (4P$_{3/2}$) the hyperfine structure of $^{40}$K is
inverted with respect to that of $^{39}$K.

This work results from the cooperation of two independent laboratories, one
in Italy at LENS, and the other in Brazil at IFSC-USP, which have
investigated the trap loss rate for the two isotopes independently. The
experimental conditions for trapping both isotopes were kept as close as
possible to allow comparison, although there were some differences,
especially in the detuning of the trapping light, needed to optimize the MOT
operation. The Brazilian experiment has performed measurements in $^{39}$K,
using an experimental setup fully described elsewhere \cite{Santos98}. In
brief, the MOT operates in a room temperature vapor cell, and the trapping
beams comes from an external stabilized Ti:Sapphire laser tuned $40%
\mathop{\rm MHz}%
$ to the red from the 4S$_{1/2}(F=2)\rightarrow $ 4P$_{3/2}(F^{\prime }=3$)
atomic transition. The repumper is provided when the carrier beam passes
through an electro optical modulator that introduces $462%
\mathop{\rm MHz}%
$ frequency shifted sidebands. Using high laser intensity ($150%
\mathop{\rm mW}%
/%
\mathop{\rm cm}%
^{-2}$) up to $5\times 10^{8}$ atoms are loaded in the MOT at a density of 10%
$^{10}%
\mathop{\rm cm}%
^{-3}$. The sudden change of intensity technique, developed by Santos \cite
{Santos98}, was used to measure the trap loss rate at low light intensities.
The MOT is loaded at full intensity and then, after the steady state has
been reached, the intensity is suddenly reduced by introducing a calibrated
neutral density filter across the trap laser beam path. After the intensity
reduction, the number of trapped atoms decreases to reach a new steady-state
value, with a temporal evolution determined by the trap loss rates in the
low intensity regime. The transient variation of atom number and density are
therefore monitored by measuring the fluorescence coming from the atoms with
a calibrated photodiode and charge-coupled device (CCD) cameras. This
technique, based on the detection of the unloading of the weak MOT, allows
to determine the loss rate also in a very low intensity regime ($I\approx
I_{S}$), differently from the standard method based on the MOT loading
analysis. Indeed, the latter method would not work at very low intensities
once the atomic density and, therefore, the collisional rates are too small.

The Italian experiment performed the measurements with $^{40}$K, loaded from
a $5\%$ enriched vapor sample. The laser cooling and trapping is carried
using the 4S$_{1/2}(F=9/2)\rightarrow $ 4P$_{3/2}(F^{\prime }=11/2)$ atomic
transition, while repumper beams are tuned to the 4S$_{1/2}(F=7/2)%
\rightarrow $ 4P$_{3/2}$($F^{\prime }=9/2$) hyperfine transition as
indicated in Fig.~\ref{levels}. Both trapping and repumping frequencies come
from a single-mode Ti:Sapphire by means of two independent acousto-optic
modulators (AOMs). A more detailed description of the MOT setup can be found
in \cite{Modugno99}, and the main MOT parameters are compared to the $^{39}$%
K ones, in Table \ref{MOTpar}. In the Italian experiment the sudden laser
intensity reduction is obtained by a quick power lowering of the AOM radio
frequency, providing the trapping beams, after a typical $4%
\mathop{\rm s}%
$ loading phase at full intensity.

As described above, to obtain the trap loss rates we measure the time
evolution of the total number of trapped atoms after the sudden decrease of
the laser intensity. The first step is to load the potassium MOT at full
intensity during a few seconds. The rate equation describing the behavior is:

\begin{equation}
\frac{dN}{dt}=L_{0}-\gamma N-\beta n_{0}N\,,  \label{rate1}
\end{equation}
where $L_{0}$ is the loading rate, $\gamma $ is the loss rate due to
collisions between the trapped K atoms and the hot background vapor, and $%
\beta $ is the loss rate resulting from collisions among the trapped
potassium atoms due to inelastic mechanisms. As first observed for Cs \cite
{Walker90} and verified for $^{39}$K \cite{Santos98}, the MOT is loaded at
constant atomic density $\left( n_{0}\right) $, so that the solution of Eq.~%
\ref{rate1} is 
\begin{equation}
N\left( t\right) =\,N_{0}\left[ 1-e^{-\left( \gamma +\beta n_{0}\right) t}%
\right] .  \label{sol1}
\end{equation}
On eqs.\ref{rate1} and \ref{sol1} above, $N_{0}$ stands as the total number
of trapped atoms achieved at the steady state, after the loading at high
intensity. The second step is to suddenly decrease the trap laser beam
intensity down to some fraction of the initial total intensity. Therefore,
the rate eq. \ref{rate1} will have to change to support the new experimental
conditions. The new equation has to link two steady-states: the initial,
containing $N_{0}$ atoms trapped at $n_{0}$ density, and the final $\left(
t\rightarrow \infty \right) $, containing $N_{1}$ trapped at $n_{1}$
density. Also, at the new intensity, the MOT\ still continues to load atoms,
but with a smaller loading rate, $L_{1}$, that has to be a fraction of the
initial one $\left( L_{0}\right) $, at full intensity. Thus, the time
evolution for the trapped potassium atoms, after the intensity decrease,
becomes:

\begin{equation}
\frac{dN}{dt}=L_{1}-\gamma N-\beta n_{1}N\text{.}  \label{rate2}
\end{equation}
The solution for the above rate eq. \ref{rate2} can be easily found, by
ordinary means, simply imposing the correct boundary conditions. That is, 
\begin{equation}
N\left( t\right) =N_{1}+\left( N_{0}-N_{1}\right) e^{-\left( \gamma +\beta
n_{1}\right) t},  \label{sol2}
\end{equation}
and it presents an exponential decay solution linking the two steady state
regimes. Note that in eq.\ref{sol2}, $N_{1}$ has to be smaller than $N_{0}$
and that is obviously the situation achieved during the experiment, once the
trap laser intensity was suddenly decreased down to a small fraction of the
total initial one.

\section{Results and Discussions}

In Fig.~\ref{exp} the experimental results for the trap loss rate $\beta $
of the two isotopes are presented as a function of the light intensity,
normalized to the saturation intensity of each isotope. Each $\beta $ value
results from the average of five independent measurements, and the error
bars represent the resulting variance. The general behavior of trap losses
for the two isotopes is similar: from high intensity down to about $%
I=15I_{s} $, $\beta $ shows a small variation with intensity, which at lower
intensities we observe an increase of trap loss that is sharper for $^{39}$K
than for $^{40}$K. This increase in trap loss rate at low intensity has been
associated to HCC. Also in this case, HCC could explain the observed
increase of $\beta $ for $^{39}$K, because the atoms are trapped in the
higher hyperfine ground state $F=2$ may change to the ground $F=1$ state
through inelastic collisions with subsequent release of the hyperfine energy.

However, for $^{40}$K this explanation for the raise of $\beta $ is not
correct, since the atoms are trapped mostly in the lowest energy ground
state $\left( F=9/2\right) $, and therefore HCC are expected to be reduced.
In fact, a quantitative estimation of the HCC contribution to the MOT losses
can be performed using the HCC rate measured in an optical trap \cite
{Roati01}. Those measurements were performed at approximately the same
temperature of the MOT, and very likely with the same distribution of
population on the magnetic substates, allowing for a fair comparison. Since
in the $^{40}$K MOT the intensity of the repumper beams remains large when
the trapping intensity is reduced, we estimate the ratio of densities in $%
F=7/2$ and $F=9/2$ states to be 
\begin{equation}
r=\displaystyle\frac{n_{7/2}}{n_{9/2}}\leq \displaystyle\frac{1}{50}\,,
\end{equation}
in the whole range of intensities explored. Thus the main contribution to
the losses due to HCC is expected to come from the ($F=9/2$, $F=7/2$) $%
\rightarrow $($F=9/2$, $F=9/2$) channel, for which we measured a rate $%
G_{7/2,9/2}=4(2)\times 10^{-12}%
\mathop{\rm cm}%
^{3}%
\mathop{\rm s}%
^{-1}$ \cite{Roati01}. Assuming HCC to be the dominant loss channel, the
time evolution of the MOT atom number is thus described by 
\begin{equation}
\frac{dN}{dt}=L-\gamma N-G_{7/2,9/2}\int n_{7/2}(r)n_{9/2}(r)d^{3}r\,,
\label{hcc}
\end{equation}
which in the case of a constant density $n$ becomes 
\begin{equation}
\frac{dN}{dt}=L-\gamma N-G_{7/2,9/2}\,r(1-r)nN\,.  \label{hccb}
\end{equation}
Comparing Eq.~\ref{hcc} with Eq.~\ref{rate2}, we get $\beta \leq 10^{-13}%
\mathop{\rm cm}%
^{3}%
\mathop{\rm s}%
^{-1}$, which is in fact between two and three orders of magnitude smaller
than the experimental observation. The possible contribution of HCC to the
losses for $^{40}$K is therefore ruled out, unless an enhancement of
collisional rate in presence of near resonant light is assumed, as proposed
for Rb in ref.\cite{Nesnidal,Gensmer}.

An alternative explanation for our experimental observations is based on the
model proposed in \cite{Telles}, which can fully explain the behavior of the
trap loss rate as in Fig.~\ref{exp} without relying on the existence of HCC.
In brief, it is possible to model the trap losses in a MOT based exclusively
on the radiative escape for any light intensity, by applying the well-known
Gallagher-Pritchard theory \cite{Gallagher89} associated with an intensity
dependent escape velocity model. According to recent observations \cite
{Bagnato00}, the escape velocity seems to follow quite well a simple model
based on the damping portion of the radiative force, that predicts a sudden
reduction of the escape velocity at low laser intensity. As already
mentioned, an interesting result of GP theory was its capability of
reproducing the experimental observations for the $\beta $ variation with
light intensity in a $^{87}$Rb MOT \cite{Nesnidal,Telles} that could not be
easily explained within the theory based on HCC.

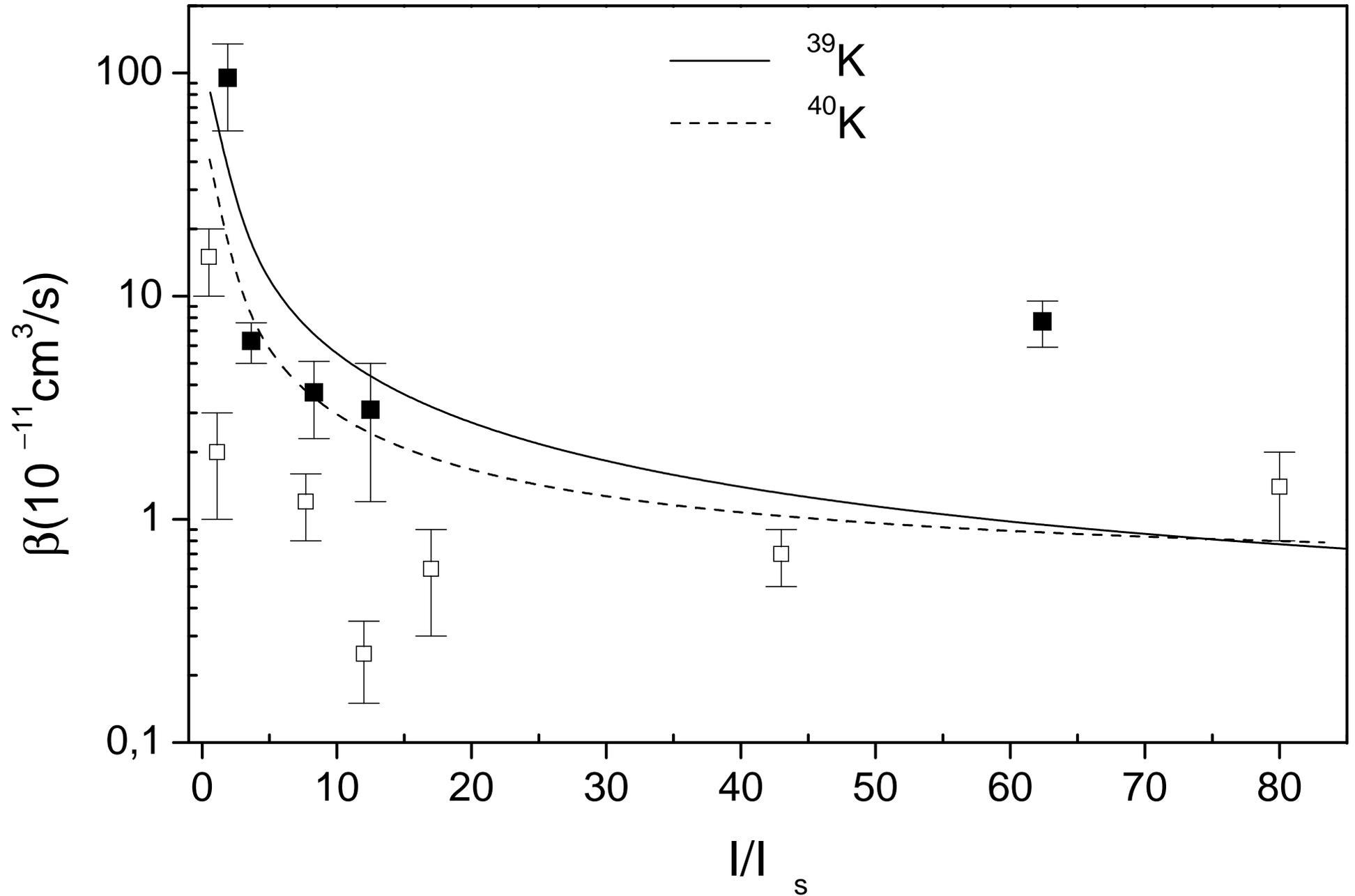
\begin{figure}[tbp]
\caption{Comparison of the experimental results and the simulation for loss
rates due to RE for the two potassium isotopes.}
\label{theo}
\end{figure}

We have therefore applied such model to our potassium MOTs, calculating the
dependence of the escape velocity on the light intensity as discussed in 
\cite{Bagnato00}. The result of the simulation for the $\beta $ variation
with light intensity is presented in Fig.~\ref{theo}, together with the
experimental results. It reproduces quite well the behavior observed in the
experiments, presenting the rise up at low intensity and an almost constant
value for high intensity for both isotopes. Therefore, the model assuming RE
as the main trap loss mechanism can reproduce all the observed features not
only for the $^{40}$K losses, but also for those in $^{39}$K, which were
expected to be dominated by HCC at low intensity. This result indicates that
RE might be in general the main loss channel in an alkali MOT.

The discrepancies between the theory and the experiment shown in Fig.~\ref
{theo} can be qualitatively explained considering the approximations present
in the model we use. In particular, for the conditions actually used in the
two MOTs, the simulation predicts a larger trap loss for $^{39}$K,
contrarily to the experimental observation. We note that our theory \cite
{Telles} assumes a two level atom, and the main parameters which determine
the loss rate are the intensity and detuning of the MOT light, together with
the $C_{3}$ coefficient of the molecular potential. For the latter quantity
we have used the value suggested in \cite{Marinescu}, while for the first
two we have used just the experimental values. The simulation predicts a
higher value of $\beta $ for $^{39}$K, since larger detunings result in a
smaller trap depth potential while compared to $^{40}$K. When considering
the multilevel character of the atoms, one should note that the $^{39}$K
trapping laser is tuned to the red of the whole excited hyperfine manifold,
due to the small separations between the states, and therefore one would
expect that the trapping effectiveness is reduced. On the other hand, an
opposite behavior for the trap losses might be expected from very general
features of the molecular potential along with RE can take place. Indeed, in
the case of $^{39}$K the trapping laser promotes coupling of the atoms to
the molecular potentials having as asymptotes the lower atomic hyperfine
components. The presence of a large number of state-mixings and avoided
crossings in this region is possibly resulting in a reduced atomic flux
towards the attractive states connecting asymptotically to the higher
hyperfine state, which mostly contributes to the losses at low light
intensities. The situation is quite different for $^{40}$K, since the
trapping laser couples the atoms directly with the attractive molecular
potentials, because of the inverted hyperfine structure of the excited
level. In this sense, the fermionic isotope is closer to the approximations
made in the model, and this seems to be the reason for the better agreement
with the theory, at least in the low intensity regime. A better agreement
between the magnitude of $\beta $ given by the theory and those observed in
the experiment would be only reach using an improved modeling, including the
multilevel atomic character. Nevertheless, we also note that the present
simulation predicts the correct intensity values in which $\beta $ suddenly
starts to increase.

\section{Conclusions}

In conclusion, we have measured and compared the trap loss rate as a
function of the MOT light intensity for two potassium isotopes, $^{39}$K and 
$^{40}$K. The measured behavior of $\beta $ shows in both cases the general
features already observed for other trapped alkali species, including the
sharp increase of the losses at low light intensities. In the case of $^{39}$%
K such increase had been previously explained with the assumption that
losses at low light intensity were dominated by hyperfine changing
collisions. To explain the data obtained for $^{40}$K we have instead to
rely in a theory where radiative escape is the dominant process for all
ranges of intensities. The dependence of trap loss rate on the laser
intensity calculated according to this model seems to fit quite well the
experimental observations for both potassium isotopes, therefore supporting
the theory of alternative trap loss channels in a MOT \cite{Telles}.

\section{Acknowledgments}

This work was carried out at Centro de Pesquisas em \'{O}ptica e
Fot\^{o}nica (CePOF) and supported by FAPESP, CNPq through the PRONEX
program and by European Community Council through Contract HPRICT1999-00111.

\begin{table}[h]
\caption{Typical conditions used for magneto-optical trapping of the
potassium isotopes.}
\label{MOTpar}
\begin{tabular}{lcc}
Isotope & $^{40}$K & $^{39}$K \\[2mm] \hline
Detuning $(%
\mathop{\rm MHz}%
)$ & $-$19 & $-$40 \\ 
HWHM of Gaussian beams $(%
\mathop{\rm mm}%
)$ & 7.5 & 6.5 \\ 
Saturation Intensity $(%
\mathop{\rm mW}%
/%
\mathop{\rm cm}%
^{2})$ & 1.8 & 1.66 \\ 
Total Number of loaded atoms & 10$^{7}$ & 10$^{8}$ \\ 
Density $(%
\mathop{\rm cm}%
^{-3})$ & 10$^{10}$ & 10$^{10}$%
\end{tabular}
\end{table}

\begin{figure}[tbp]
\caption{Experimental loss coefficient for the potassium isotopes $^{39}$K
(open squares) and $^{40}$K (black squares).}
\label{exp}
\end{figure}

\end{document}